# Causality in Static Models as an Initial Constraint on the Chronology of Events in System Behavior

Sabah Al-Fedaghi

Computer Engineering Department
Kuwait University
Kuwait
sabah.alfedaghi@ku.edu.kw

*Abstract*—This paper analyzes the notion of causality in a conceptual model, mainly as applied in software engineering. Conceptual system modeling can be considered a three-level process that begins with building a static structural description to develop a dynamic model that will identify events used to specify the chronology of events. In this context, the model involves a representation of a portion of reality, based on ontology of different kinds of things and their basic relations to each other. Relations are defined in terms of their participating entities. This paper concerns relations between events, specifically causal relations among events in modeling. We examine causality in many fields of study to understand its role in modeling. The problem is that, according to many researchers, causation is so inextricably bound up in misleading associations that it is hard to define and is shrouded in mystery, controversy, and caution. We study and clarify the notion of causality through several examples, utilizing an event definition as a time thing/machine in a new conceptual modeling methodology. In conclusion, we claim that the purpose of causal relations in a system's static description is to constrain the system's behavior and thus exclude some possible chronologies of events.

*Keywords-causality; static model, behavioral model; conceptual representation; events; cause and effect*

## I. INTRODUCTION

This paper analyzes the notion of causality in a conceptual model, mainly as applied in software engineering. The purpose of specifying causality in a system's static description is to constrain the system's behavior and thus exclude some possible chronologies of events. "Specification" here refers to descriptive modeling (e.g., in contrast to mathematical) aimed at abstractly describing the actual state of affairs in a system as a well-founded basis for a cyberphysical system's design and implementation. The model also provides comprehensibility of the modeled portion of real-world aspects. Such a development involves gradual evolution at different levels of abstraction, progressing through static (e.g., classifying objects), dynamic (e.g., identifying events), and behavioral (e.g., specifying the chronology of events) phases.

In this context, a model involves a representation of the modeled potion of reality, including the existence of different kinds of things and their basic relations with each other. Relations are defined in terms of their participating entities, and according to Guizzardi, Herre, and Wagner [1], "Without relations the world would fall into many isolated pieces." This paper concerns relations among events, specifically causal relations among events in modeling.

### A. Importance

The ability to learn about the notion of causality is a significant component of human-level intelligence and can serve as the foundation of artificial intelligence [2]. All sciences seem to involve studying the same things: matter, causalities, and time [3]. The concept of causation seems to be extremely important, "though, as we shall see, this claim has been disputed" [4]. An enormous amount of literature exists on causality, especially in artificial intelligence, philosophy, economics, statistics, genetics, artificial intelligence, and other disciplines.

### B. Problem with Causality

Russell complained that the word "cause" is "inextricably bound up with misleading associations" [4]. According to Pearl [5], "Though it is basic to human thought, causality is a notion shrouded in mystery, controversy, and caution, because scientists and philosophers have had difficulties defining when one event truly causes another." Causality is hard to define, and we often only intuitively know about causes and effects [6].

### C. What Is Causality?

Plato in *Timaeus* stated that "everything that becomes or changes must do so owing to some *cause*, for nothing can come to be without a cause" [7] (italics added). According to Aristotle [7], the response to the question "What is this?" could be answered in one of the following ways: "This is marble," "This is what was made by Phydias," "This is something to be put in the temple of Apollo," and "This is Apollo." These responses answer four different questions, respectively, "What is this made of?" "Who is this made by?" "What is this made for?" and "What is it that makes this what it is and not something else?" These answers and questions correspond to Aristotle's four causes:

- The material cause (e.g., the marble, which can be seen as causing the existence of the statue).
- The efficient cause (e.g., Phydias, who causes the statue to exist).
- The final cause (e.g., the sake of [worshipping] Apollo for which the statue was created).

- The formal cause (e.g., the form or structure of the statue to cause its existence).

In this sense, cause is "something without which the thing would not be [and] Aristotle argued that the most important and decisive cause was the formal cause" [7].

Descartes (1596-1650) limited his examination of causation to the efficient causes of things [7]. Hobbes (1588-1679) defined causes in terms of motion, with causation being a relation between the motions of different bodies. The material and efficient causes are both part of the entire cause. As Hulswit [7] stated, "Nothing would happen if nothing moved, and the only things that move are bodies. Moreover, all causation occurs by contact—that is, it consists if the motion of contiguous bodies" (see sources in [7]). Leibniz (1646–1716) declared, "there is nothing without a reason, or no effect without a cause… Each monad develops in synchrony with all other monads. Just as a good clockmaker constructs a number of clocks that keep perfect time, so efficient causality and final causality complementary" (see sources in [7]). David Hume (1711-1776) viewed causation as "the only relation that allows us to go beyond what is immediately present to the senses, to discover either the real existence or the relations of objects" (see [4]). According to Newton (1642–1727), two classes of events exist: those that happen according to a law and those that are the effects of causes. Causation and law like behavior are mutually exclusive notions [7].

According to Beebee [4], the concept of causation has "at least in the last half-century or so pervaded philosophical theorizing." The question of its nature has "dominated the extensive literature on the topic of causation in analytic philosophy in the last fifty years or so" [4].

Causation has been explained in terms of counterfactual conditionals of the form "If *A* had not occurred, *C* would not have occurred" [8]. According to Lewis [9], "We think of a cause as something that makes a difference, and the difference it makes must be a difference from what would have happened without it. Had it been absent, its effects would have been absent as well." Lewis's [9] events are classes of possible spatiotemporal regions.

Another approach to causation not only views causality with regard to particular events or happenings but also makes general causal claims (e.g., smoking causes cancer). Such claims are rarely arrived at by generalizing over specific cases of "particular" causation but instead mainly based on statistical methods. Probabilistic theories of causation seek to analyze the notion of general causation by appealing to statistical fact [4]. Other researchers claim that in "E because C," the most basic form of a causal claim, "C" and "E" *state facts* (see [4] referencing [10]). In this paper, we are not concerned with this probabilistic type of causation.

Many definitions of event-based causality currently exist. The following is a sample list of some of these definitions. The aim is to demonstrate the diversity of this notion.

- Causation is a relationship that holds between events, properties, variables, or states of affairs [3].
- A cause or event can be one of the following:
  1. A state that persists and does not change over a period of time.
  2. An occurrence that can be subcategorized into an event, such as

- An occurrence with a culmination, or a process.
- An occurrence that is homogenous and does not have a climax or an anticipated result (see Khoo, Chan, and Niu [11] referencing Terenziani and Torasso [12]).

- Events and processes can be categorized into those that have duration (occur over a period of time) and those that are momentary. Thus, an event can be one of the following:
  1. A punctual occurrence or achievement.
  2. A punctual or momentary process.
  3. A durative process or activity that continues over a period of time (see Khoo, Chan, and Niu [11]).

- An event is a region of space-time that has certain features essentially and some only contingently. "For example, in the region I am currently occupying, several events are currently occurring, one of which is essentially my drinking coffee, another is essentially my drinking coffee slowly, and another is essentially my drinking coffee out of a mug" (see [4] referencing Lewis [9]).

### D.   In This Paper

The next section reviews, with some enhancement of the ontological theory of a thinging machine (TM). This involves a single entity that has a dual being as a *thi*ng and a *ma*chine (thimac). The term "thinging" relies on Heidegger's [13] notion of "things" [14-23]. Based on thimacs, conceptual system modeling can be considered a three-level process that begins with building a static description of thimacs to develop a dynamic model that may identify events used to specify behavior Section 3 applies TM to Aristotle's four causes. Sections 4- 7 apply the TM methodology to different types of modeling.

## II.    THINGING MACHINE

The TM ontology is based on a single category called thimacs. A thimac crystallizes being into dynamic forms and kneads together an "object" (called a *thing*) and a "process" (called a *machine*)—thus, the name thimac. The thimac notion is not new. In physics, subatomic entities must be regarded as particles and waves to fully describe and explain observed phenomena [24]. According to Sfard [25], abstract notions can be conceived in two fundamentally different ways: structurally as objects/things (static constructs) and operationally as processes. Thus, distinguishing between form and content and between process and object is popular, but processes and objects must unite in order to understand TM bases modeling.

A TM adopts this notion of duality in conceptual modeling, by generalizing it beyond mathematics and its utilization in software engineering modeling. "Structural conception" means seeing a notion as an entity with a recognizable internal structure. The operational way of thinking emphasizes the dynamic process of performing actions. A model describes a given domain, independent of technological choices that could impact the implementation of a system based on itself.

## A. The Machine

The term "machine" refers to a special abstract machine called a "thinging machine" (see Fig. 1). A TM is built under the postulation that the machine only performs five generic processes: creating, processing (changing), releasing, transferring, and receiving.

A thimac has a dual being as a thing and a machine. A thing is created, processed, released, transferred, and/or received, whereas a machine creates, processes, releases, transfers, and/or receives things. We will alternate among the terms "thimac", "thing", and "machine" according to the term we want to stress at each instance.

The five TM flow operations (also called stages) form the foundation for thimacs. Across the five stages, the flow (a solid arrow in Fig. 1) of a thing means its "motion" of occupying different stages. The TM diagram reflects the succession imposed on this "motion" of the thing. The flow across the five stages is performed in accordance (agreement) with this structure. Even though this succession of TM stages is imposed by the thimac's structure, we can incorporate time by saying that the flow is the occupation of different stages at different times. Note that this definition was inspired by Russell's definition of motion as occupying different places at different times [26]. The thing in a TM has no other places beside the five generic stages. When we adopt this theory (used to solve Zeno's paradoxes [26]), the arrows in Fig. 1 have no corresponding events (time), as they do not denote transitions.

The generic flow operations of a TM can be described as follows.

- *Arrival*: A thing reaches a new machine.
- *Acceptance*: A thing is permitted to enter the machine. If arriving things are always accepted, then arrival and acceptance can be combined into the "receive" stage. For simplicity, this paper's examples assume a receive stage.
- *Processing* (change): A thing that changes without creating a new thing undergoes transformation. The terms "worked" (thing) and "work on/out" may express the sense of the process here.
- *Release*: A thing is marked as ready to be transferred outside of the machine.
- *Transference*: A thing is transported to somewhere outside of the machine.
- *Creation*: A new thing is born (appears/stands out) in a machine, which creates in the sense that it "finds/originates" a thing: it brings a thing into the system and then becomes aware of it. Creation can designate "bringing into existence" in the system because what exists is what is found. Appearance indicates existence, but existence may not imply appearance. Creation also refers to producing, making, forming, and bringing forth. Making implies processing something to create something else.

In addition, the TM model includes memory accessed from all stages and, triggering that embraces the notion of causation.

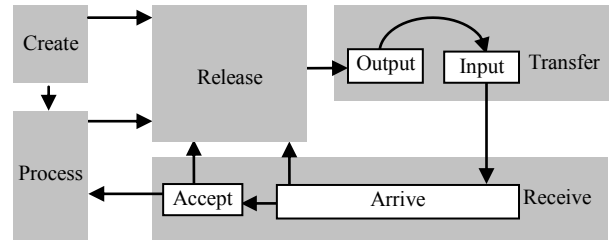

**Figure 1. A thinging machine.**

## B. Thinging Machine Language

The TM in Fig. 1 can be specified in a two-dimensional language (which we call TM language), wherein the arrows are represented by dots. For example, the following shows different flows in Fig. 1 in this TM language:

*Flow.Create.release.transfer.output*
*Flow.Create.process.release.transfer.output*
*Flow.Transfer.input.receive.arrive.release.transfer.output*
*Flow.Transfer.input.receive.arrive.accept.release.transfer. output*
*Flow.Transfer.input.receive.arrive.accept.process.release.t ransfer.output*

The dot "." is used to donate flow or containment. We will use "-->" to indicate triggering.

## C. Example

Consider Fig. 2 of a water thimac, in which a machine processes oxygen and hydrogen to create the thing water. The creation on the machine side corresponds to manifestation on the thing side. Thus, the water themac has two faces: (to create —verb, manifestation —noun). Fig. 3 illustrates an analogy from the arts of simultaneously constructing a thing from two halves.

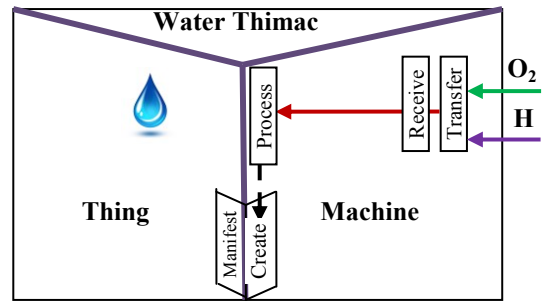

**Fig. 2. The *thi* of water exists through creation by its *mac*.**

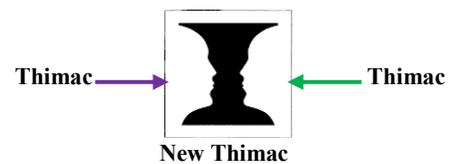

**Fig. 3. An image that illustrates the simultaneous emergence of two halves.**

Note that this study is a modeling analysis of a limited domain (i.e., an accounting system in an organization) and not a new ontological theory in philosophy. Within such a context, "manifestation" refers to an appearance in the system that is modeled. The machine is the active subject (agent) that creates, processes (changes), releases, transfers, and receives, while the thing is the inert object (patient) being subjected to all those processes. Thus, having been created by a machine implies that the thing manifests itself. The machine stands under the thing; it is the duality of being that describes the kernel of the thing. It is also the *cause* of the thing's manifestation.

Creation arises within a thimac, as a machine facilitates creation, and that creation *manifests* a thing. The thimac includes in it both a material base and an efficient cause. This means that the machine can *cause* the thing's creation. In this case, causation means processing something to bring something else into "existence", i.e., to appear in the system. In the water example, the creation as a machine operation is the "existence"/*cause* of the thing water. Fig. 4 shows water as a *mac* and a *thi*.

## III. ARISTOTLE'S FOUR CAUSES AND BACKWARD CAUSATION

As demonstrated in the Apollo example, Aristotle's four causes can occur at the same time. Fig. 5 shows a static TM representation of the four causes. In response to the question "What is this?" we say that this is a marble (1) that flows to Phydias (2), who has Apollo's image/form (3) and processes (4) the marble according to that form to create (5) a statue that is erected in the temple for worship of Apollo (6). Note that Phydias, Apollo, the workshop and the temple "exist in the model" (i.e., there is a creation stage), but the creation stage is not shown, under the assumption that the surrounding box is sufficient to imply this.

Fig. 5 can be written in TM language as follows:

*Flow.Marble.create.release.transfer.Phydias.transfer.receive.transfer.workshop.*

*Flow.Apollo.image.create.release.transfer.Phydias.transfer.receive.release.transfer.workshop.*

*Flow.Workshop.transfer.reveive.process-->Apollo.temple.statue.create.process.*

The mapping of these flows to the diagram is obvious.

To develop the corresponding dynamic model, we need to define the notion of an event. An event in a TM is a thimac with a time subthimac. For example, the event *The marble is transported to Phydia* is modeled as shown in Fig. 6. It contains three thimacs: the time, the region, and the event itself. The region is the subdiagram of Fig. 5 where the event occurs. The events may include other subthimacs (e.g., intensity). For the sake of simplicity, we will represent events by their regions if the events do not conflict.

Fig. 7 shows a dynamic model in which events are identified according to Aristotle's four causes.

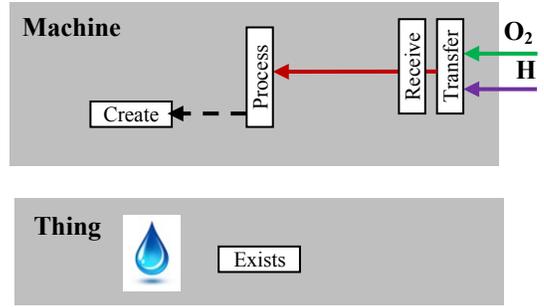

**Fig. 4. Water as a *mac* and a *thi*.**

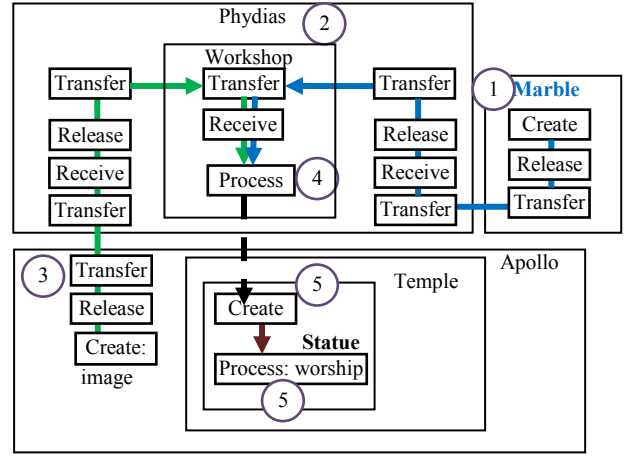

**Fig. 5. The static model of the Apollo statue example.**

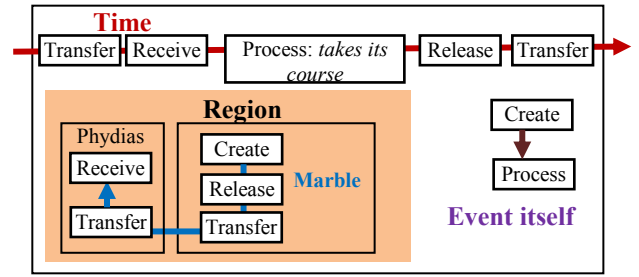

**Fig. 6. The event "The marble is transported to Phydias."**

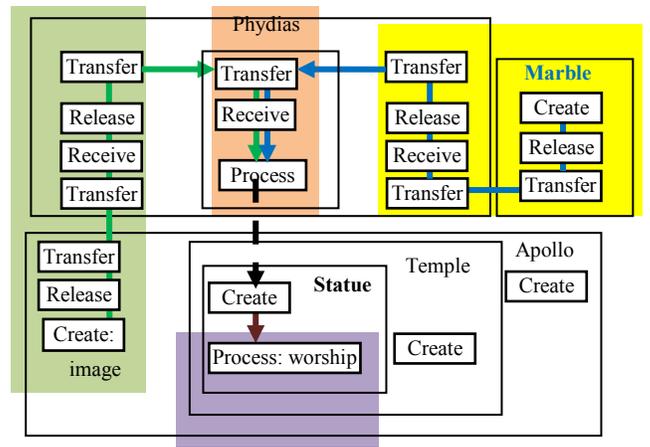

**Fig. 7. The dynamic model of the Apollo statue example.**

A disturbing observation emerges when we construct the system's behavior in Fig. 8. The final cause seems to be an event that occurs after creating the statue. According to Khoo, Chan, and Nui [11], the final (or teleological) cause is a special type that occurs after the effect—a kind of backward causation.

For example, when a doctor prescribes a drug to treat a disease, this is because of the doctor's belief that the drug will cure the disease. An intended future effect thus causes an event in the present. This is related to mental causation since the cause stems from thought processes in the doctor's mind. Teleological cause is also seen as occurring in nature and can be understood in terms of natural selection. For example, a bird's wings can be said to exist so that the bird may fly.

According to the *Stanford Encyclopedia of Philosophy* [27], backward causation means that the temporal order of cause and effect is a mere contingent feature, and cases may exist in which the cause casually precedes its effect.

In our example, the backward causation seems to be the result of mixing up the static and behavior models. Let us start again at the static model of the Apollo statue example.

- "Worshipping Apollo through the statue" does not involve time.
- When we construct the dynamic model, subdiagrams in the model become events, including the subdiagram of the model: "worshipping Apollo through the statue". Accordingly, the *aim* (as one of Aristotle's causes) of worshipping Apollo through the statue is not in the dynamic model.
- However, the *event* of worshipping Apollo through the statue (actual worship) is in the dynamic model. For example, if the statue is created at 9:00 AM, the worship event happens at 9:10 AM.
- This implies that the aim is different from the event of actually practicing worship of Apollo through the statue. The aim is a snapshot of what to do at some point in the future. Generating such a snapshot at a certain time is an event.
- The declaration or assertion of the aim of worshiping Apollo through the statue occurs, say, one year prior to the occurrence of the event.

The static model can now be corrected to represent the two events (declaring the aim and actual worshiping), as shown in Fig. 9. The two events, which coincide in region (subdiagram), are (a) asserting the aim to worship Apollo through the statue and (b) actually practicing that worship.

Fig. 10 shows the dynamic model that contains these two events. The same subdiagram is drawn twice because it is difficult to draw a diagram over itself. Fig. 11 shows the correct chronology of events. Note that the material, sufficient, and formal descriptions in the static model can be viewed as objectives of the aim (the specific steps that can be taken to achieve the aim).

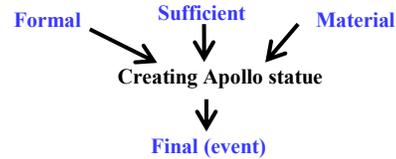

Fig. 8. The behavioral model of the Apollo statue example.

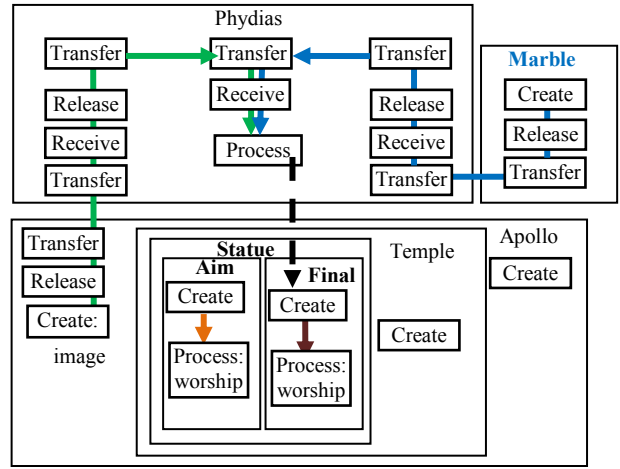

Fig. 9. The correct static model of the Apollo statue example.

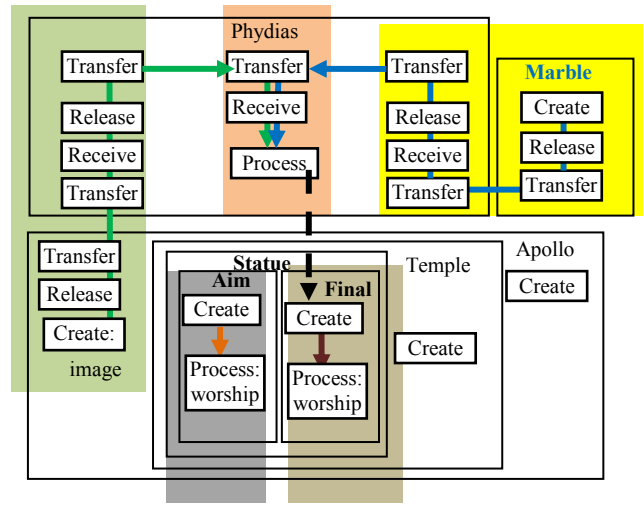

Fig. 10. The correct dynamic model of the Apollo statue example.

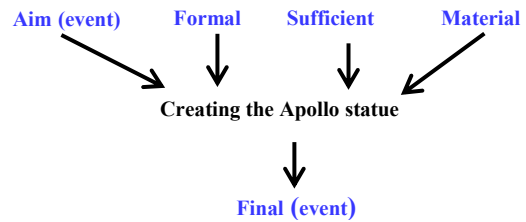

Fig. 11. The correct behavioral model of the Apollo statue example.

## IV. ROBOT PROBLEM

Consider the robot problem described by Pearl [5]: "How should a robot acquire causal information through interaction with its environment? How should robot process causal information received from its creator-programmer?" For example, assume we teach our robot all we know about cause and effect in a room. When the robot is given the information "If the grass is wet, then it rained" and "If we break this bottle, the grass will get wet," then the computer will conclude "If we break this bottle, then it will have rained" (see source in Pearl [5]). The reasoning goes as follows:

1. "If the grass is wet, then it rained."
2. "If we break this bottle, the grass will get wet."
3. "If we break this bottle, then it rained."

According to Pearl [2], "Only a few decades ago scientists were unable to write down a mathematical equation for the obvious fact that 'mud does not cause rain.' Even today, only the top echelon of the scientific community can write such an equation and formally distinguish 'mud causes rain' from 'rain causes mud.'"

Fig. 12 shows the static TM model of this grass situation. The rain falls (1) on the grass (2) to change its condition to be wet (3). The bottle breaks (4), causing the water (or any liquid) inside to fall and make the grass wet (5). Fig. 13 shows the set of events according to "it rained", "the grass is wet", and "we break this bottle".

The problem seems to mix "it rained", "the grass is wet", and "we break this bottle" as events. It is true that "The grass is wet" happens after the other two events. Accordingly, we have the acceptable behaviors does not include the two sequence, $E_1 \rightarrow E_2 \rightarrow E_3$ or $E_2 \rightarrow E_1 \rightarrow E_3$, e.g., the wetting should happens immediately after raining or breaking the glass. The acceptable behaviors are (See Fig. 14): $(E_1, E_2) \rightarrow E_3$, i.e., simultaneous rain and breaking, $E_1 \rightarrow E_3 \rightarrow E_2$, $E_1 \rightarrow E_3$, $E_2 \rightarrow E_3 \rightarrow E_1$ and $E_2 \rightarrow E_3$. Accordingly, event $E_2$ ("we break this bottle") never implies the event $E_1$ ("it rained") because after $E_2$ the event $E_1$ may or may not happens.

## V. EQUATIONS AND EVENTS

According to Pearl [5], a circuit diagram shows cause–effect relations among the signals in a circuit. Switching from logical gates to linear equations, Pearl [5] assumed a system of linear equations with the following two equations:

$$Y = 2X, Z = Y + 1, \text{ and}$$

$$X = Y/2, Y = Z - 1.$$

The issue for Pearl [5] was that these equations were equivalent to their circuit diagrams, and according to Pearl, "The top one tells us that if we physically manipulate Y it will affect Z, while the bottom one shows that manipulating Y will affect X and will have no effect on Z." Pearl [5] then discussed the effect of physically manipulating Y—for example, by setting Y to 0. His conclusion is that from this pair of equations alone, "it is impossible to predict the result of setting Y to 0, because we do not know what surgery to perform – there is no such thing as "the equation for Y.""

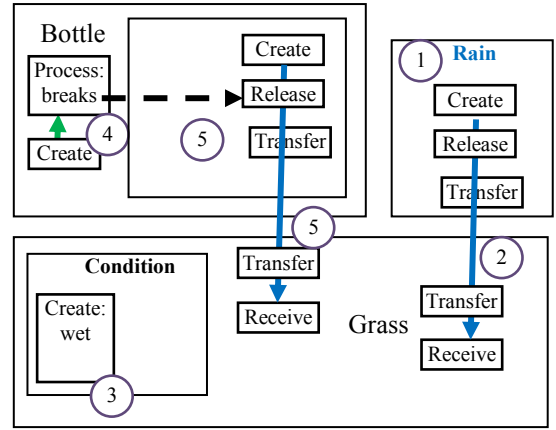

**Fig. 12. Static model of the grass example.**

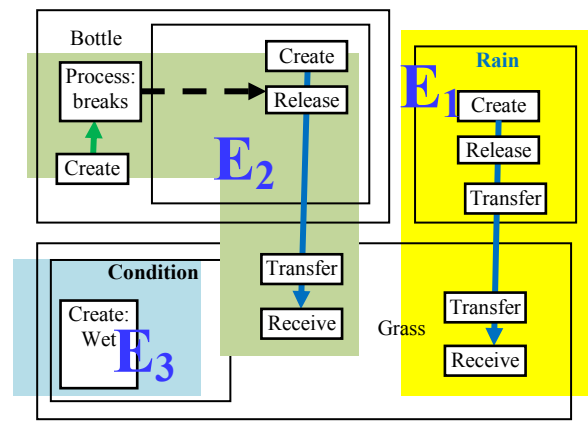

**Fig. 13. Dynamic model of the grass example.**

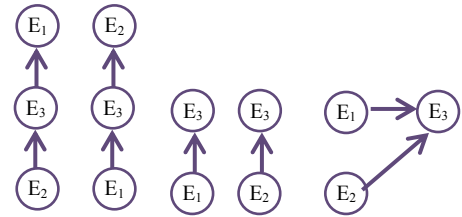

**Fig. 14. All acceptable behaviors of the wet-grass example.**

Examining this issue using the TM model shows that this system of equations mixes up the notions of operations and events. The equations involve operations such as multiplication and addition. Fig. 15 shows the static description of this system of equations as follows:

"$Y = 2X$" is modeled by the flow of X (circle 1) and 2 (circle 3), to be multiplied (circle 4) to produce Y (circle 4).

"$Z = Y + 1$" is modeled by the flow of Y (circle 5) and 1 (circle 6), to be added (circle 7) to generate Z (8).

"$X = Y/2$" is modeled by the flow of Y (circle 9) and 2 (circle 10), to be divided (circle 11) to produce X (circle 12).

"$Y = Z - 1$" is modeled by the flow of Z (circle 13) and 1 (circle 14), to be subtracted (circle 15) to produce Y (circle 16).

As previously stated, the static model contains all things and machines in the system at all times.

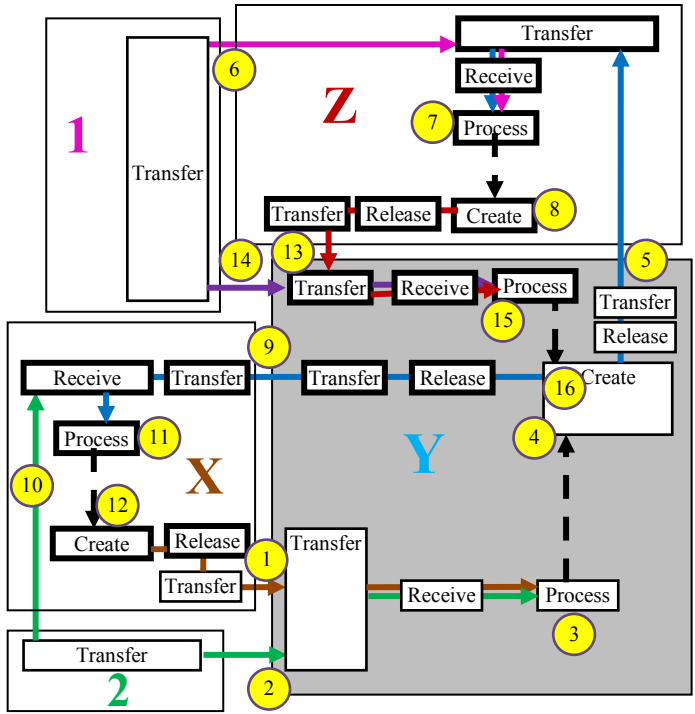

**Fig. 15. The static model of the system of equations.**

Fig. 16 shows the dynamic model. The events are identified and named according to the operation being performed, in large black characters. Fig. 17 shows the behavior of the set of equations {Y = 2X, Z = Y + 1}. In Fig. 17, X is the starting point of the multiplying event to create 2X, then the event of finding the value of Y, and so on. This means that Y cannot be found without 2X previously being calculated, but X should have a value before 2X is calculated, and so on.

Fig. 18 shows the behavior of the set {X = Y/2, Y = Z − 1}. Y is first given a value so that Y/2 (circle 11) can be calculated, and then X has the value of the calculation in Y/2 (12). These events can precede 2X but never follow it by Y (see Fig. 18), *because Y is an event that has already happened*. Accordingly, the two sets of events are not equivalent in terms of their chronology, just as the events sleeping→eating→dressing→work is not "equivalent" to the events sleeping→eating→dressing→eating→work.

The issue here is the chronology of events, not the manipulation of variables or equations. This mix-up appears when Pearl [5] examines the effects of setting Y to 0. Note that the two behaviors in Fig. 17 do not permit Y = 0 as an additional equation because the equation starts with X and that force puts 2X in Y (Y is not an independent variable). According to the behaviors in Fig. 17, the only way to set Y to 0 is to set X to 0, find 2X, and then put the result into Y. In Fig. 18, setting Y to 0 makes X = 0.

Assume that Y = 0 is added to the system (zero flows to Y, similarly to the flows of 1 and 2 in the static model). In this case, we have a third behavior, shown in Fig. 19. Thus, no mysterious surgical process is involved here; it is an issue of overlapping events.

In a TM, we can say that the cause of what happens in event Z is the event Y + 1, which in turn is caused by Y, which is caused by 2X, which is caused by X. Thus, the cause is the chronology of events that led to Z. It is impossible to predict the result of setting Y to 0 because no behavior in the system permits doing so. We handle this situation by specifying a third behavior that starts at Y. The chronology of events is an important notion in programming. Thus (x + y) − z specifies the event (x + y) before the event (resultant sum + z).

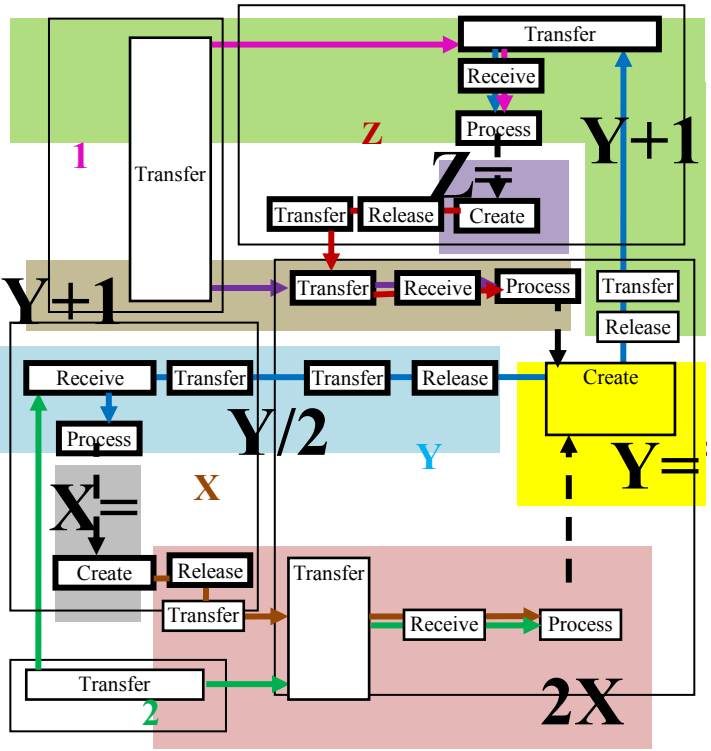

**Fig. 16. The dynamic model of the system of equations.**

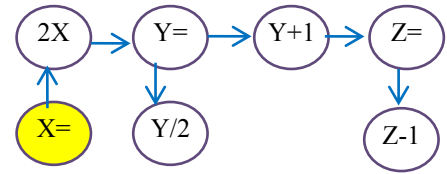

**Fig. 17. The dynamic model of {Y = 2X, Z= Y+1}.**

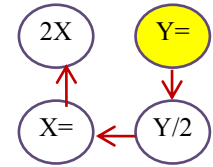

**Fig. 18. The dynamic model of {X = Y/2, Y = Z-1).**

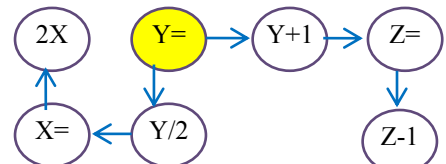

**Fig. 19. The behavior that include Y=0.**

## VI. SEMANTICS OF TRIGGERING

Because we now have a better understanding of the semantics of causation, this section addresses how a TM diagram incorporates the notion of causality. This leads us to conclude that causality in static models is an initial constraint on the chronology of events in system behavior.

The flow in a TM (the solid arrows in the diagrams) forces a chronology of events. If there is an arrow from a stage A to a stage B in the static description, then event B happens after event A. This interpretation can be generalized to non-basic events: if there is flow between events, then their sequence in the behavior model is decided. For example, if there is a flow from Y to Y+1 in Fig. 16, then event Y precedes event Y+1. If there are two independent (with respect to flow) subdiagrams, then we can ask, what is the acceptable sequence of their events? It is possible to specify (a) a certain sequence of events or (b) let the events loose as different starts for acceptable behavior. These and other cases justify introducing triggers into the TM diagram. Triggering specifies logical sequencing at the static level. For example, the creation of water *results from* the *processing* of oxygen and hydrogen (Fig. 2). If we do not include a dashed arrow, then the event of creation may occur regardless of the occurrence of the oxygen and hydrogen being processed.

One of the mechanisms facilitated by triggering is causation. For example, Pearl [5] wrote, "What difference would it make if I told you that the rooster does cause the sun to rise? The obvious answer is that knowing what causes what makes a significant difference in how people act." From a TM perspective, knowing causation is a way to exclude certain behaviors. There is no flow from the rooster to the sun; that is, the sound does not flow to the sun. If we draw the behavior of the {rooster, sun} system, then we have two events:

$E_1$: The event of the sun rising
$E_2$: The event of the rooster crowing

Assuming that $E_2 \rightarrow E_1$, this is the result of the static model: *Rooster.sound.create.release.transfer.sun.transfer.receive.process-->Rising.create.*

In this case, a person who says that the rooster causes the sun to rise must show that the rooster's sound travels at about 150 million km. Because this is scientifically false, the speaker model has to be as follows:

*Rooster.sound.create-->Sun.rising.create.*

The speaker has to use the dashed arrow "-->" (indicating triggering) in the model because the thing sound is ontologically different from the thing rising (noun). The issue here is dissimilarity in models; Hume (1711-1776) asked how we might know that a flame has caused heat, arguing that the causality of the events rested on an untestable metaphysical assumption. In TM language,

*Flame.create-->Flame.heat.create.*

It is not the case that

*Flame.create-->Heat.create.*

Heat, in this sense, is just a property of the flame.

In the context of modeling, using a dashed arrow in a TM is justified by the agreement of all participants in the system: the users, designers, implementers, and so on.

## VII. MODELING AN ELEVATOR

This section provides an example of a model that contains a great deal of triggering.

Hoss [28] modeled an elevator system in a case study, using 16 textual use cases, use case diagrams, sequence diagrams, class diagrams, and state machine diagrams. Figs. 20 and 21 show a sample use case and a sequence diagram, respectively.

The static and dynamic TM models of an elevator system are shown in Figs. 22 and 23, respectively, assuming one elevator. The static model can be described as follows:

### A. Static Model

The passenger presses (circle 1) a button (2) to trigger (3) the creation of a signal requesting the elevator (4). The arrival of the signal (6) triggers information being sent to the controller about the passenger floor from where the signal originated. The signal flows (7) to the controller (8), where it is processed (9) to generate a light signal (10), which flows to the system on the passenger floor (11) and then to the button (12), turning the button light on (13).

The passenger floor (14) and the floor where the elevator is located are compared (16), and the result depends on this comparison.

### A.1 Elevator on the Passenger's Floor

If the elevator is currently on the passenger floor (17), then a signal is created (18) and sent (19) to the door on the passenger floor to open it (20). Simultaneously,

- A signal is created (21) to turn the button light off (22).
- The time of the system clock (23) is registered to count a certain period of time before closing the door (24).

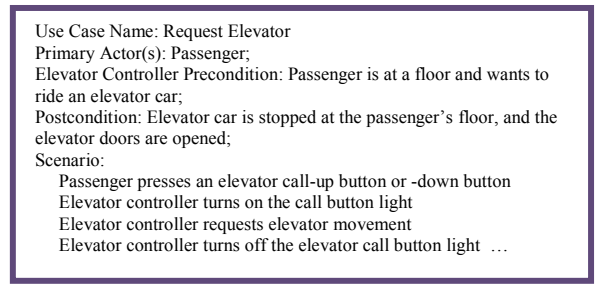

**Fig. 20. Sample use case for the elevator system (redrawn, partial from [28]).**

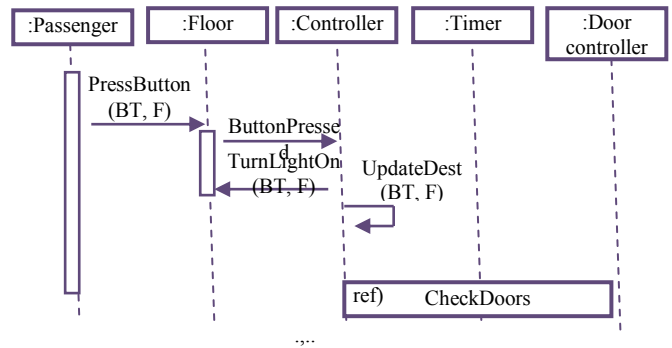

**Fig. 21. Passenger request sequence diagram (redrawn, partial from [28]).**

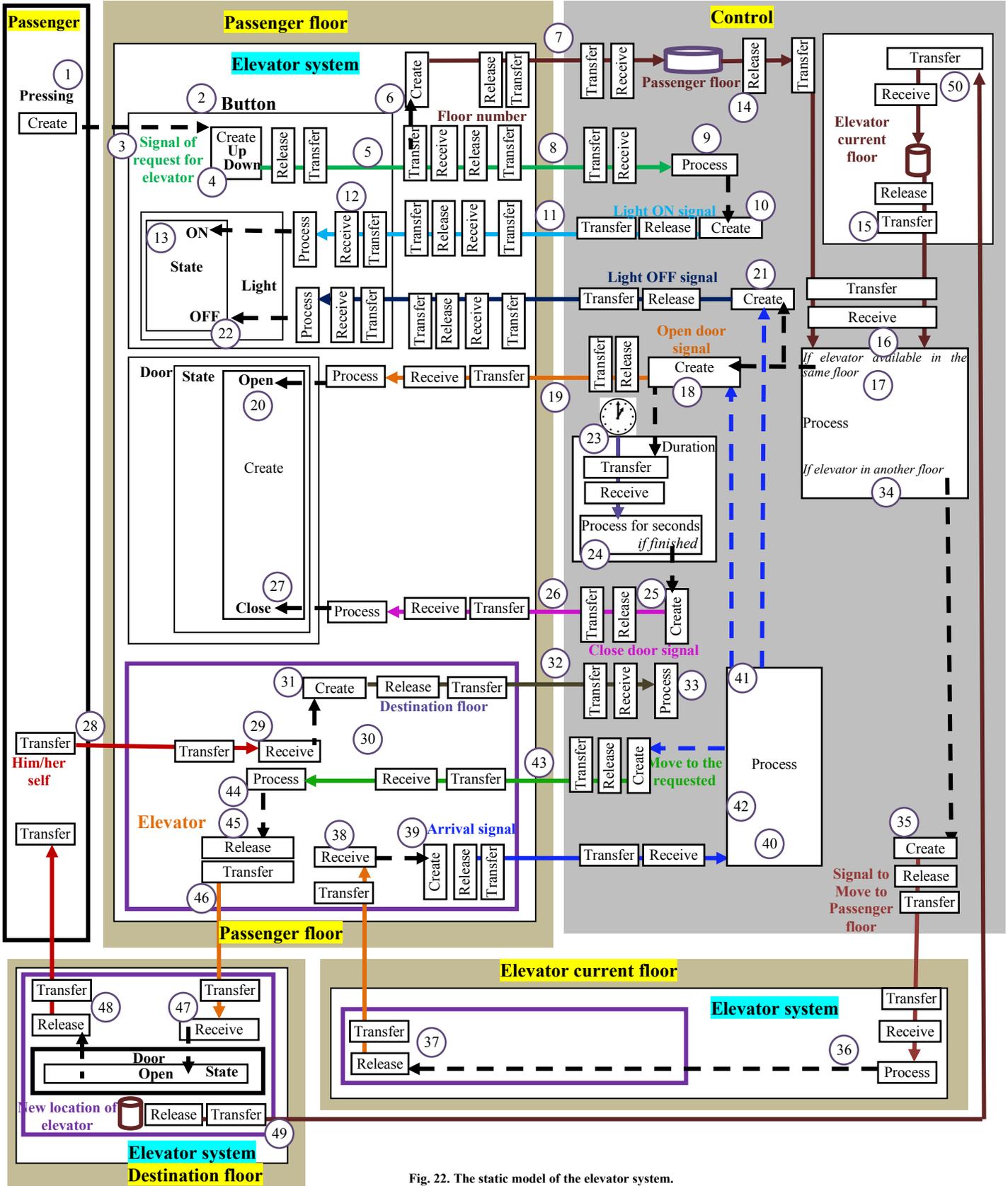

**Fig. 22. The static model of the elevator system.**

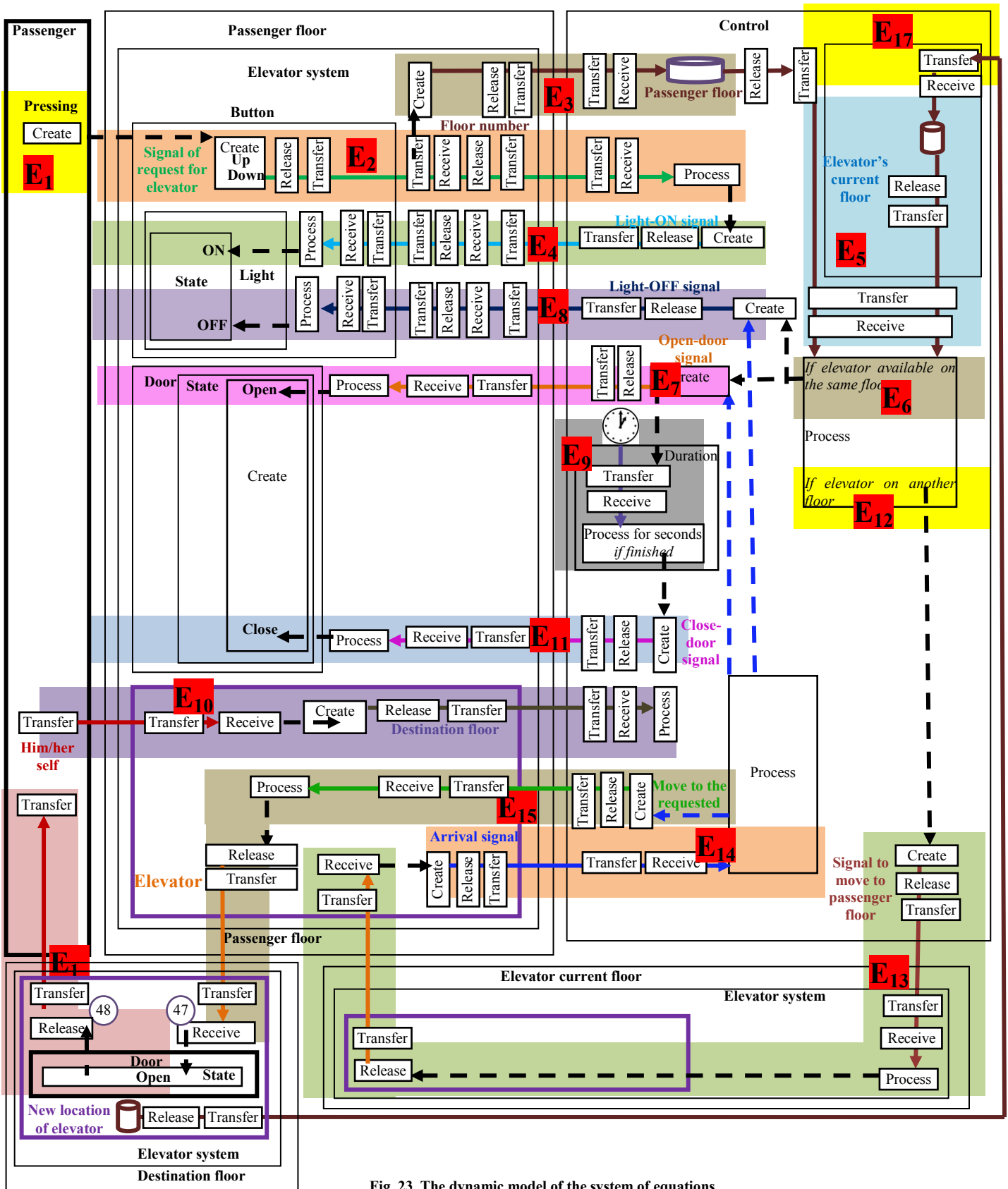

Fig. 23. The dynamic model of the system of equations.

Upon the wait time elapsing, a signal is created (25) and sent (26) to the door to close it (27).

In the period between the door's opening and closing, the customer enters (28) the elevator (29 and 30) and then selects the destination floor (31), which flows to the controller (32) to be processed (33).

### A.2 Elevator on a Different Floor Than the Passenger Floor

If the elevator is located on a different floor (34), a signal is created (35) to trigger (36) the elevator's movement (37) to the passenger's floor (38).

The elevator's arrival to the passenger's floor (38) triggers the creation of a notice of this arrival (39).

The notice is processed (40) to trigger (41) the steps in *A. Elevator on the Passenger's Floor*.

The notice is processed (40) to trigger (42) the creation of a signal (including the destination floor), which flows to the elevator (43).

The processing (44) triggers (45) the elevator's movement from the passenger's floor to the destination floor (46 and 47).

The door opens, and the passenger leaves the elevator (48). The location of the elevator is updated (40 and 50).

### B. Dynamic Model

In the dynamic model, shown in Fig. 23, we observe the following:

- The sequence $E_1 \rightarrow E_2$ is defined by triggering in the static model because a direct flow is impossible between the physical pressing and the signaling of the need for the elevator. The modeler adds triggering to indicate that a causality relation exists between pressing a button and generating a signal. Otherwise, the system would permit pressing without a signal and/or signaling without pressing.
- $E_2$ is made of basic events, whose sequence is defined by the flow of the requesting signal.
- The modeler creates $E_2 \rightarrow E_3$ to make $E_2$ cause $E_3$. Such a triggering process starts another flow of a different thing.

The triggering mechanism is like a domino effect, propagating the event of a piece falling onto another event piece in a dynamic cascade. The player has to set the right distance between two pieces to ensure that the second event will follow the first; otherwise, the first piece may or may not hit the second piece. Similarly, in the static model, triggering guarantees that $E_3$ happens after $E_2$.

- The event sequence $E_3 \rightarrow E_5$ is defined by the flow of the passenger floor (number) to event $E_5$, to be compared to the elevator flow. This is not a subjective sequence of events. The modeler has to bring the two flows together to compare them, which is performed after the data arrive.
- $E_3 \rightarrow E_5$ does not involve any relation between the two events. Opening the elevator door may or may not be followed by the customer entering the elevator. Yet, if he/she enters the elevator, the door surely will be open. There is no triggering in this case, and the modeler decides the sequence of events.

We can continue examining every relation between events to construct Fig. 24, which shows the elevator system's behavior. Accordingly, we can identify the different relations between the two events in Fig. 24 to decide whether the behavior is based on flow (black arrows) or something else (red arrows), as shown in Fig. 25.

### VIII. CONCLUSION

In this paper, we have reviewed the notion of causality to explain its role in modeling. Our analysis leads us to conclude that casual relations are identified by the modeler (subjective), in contrast to flow relations, which comprise their own sequences of happenings.

The modeler is analogous to a weaver who interlaces the flow and basic five stages of a TM, creating a form out of the fabric. He/she is also like a tailor who sews various fabrics together (triggering) to produce different forms (behavior).

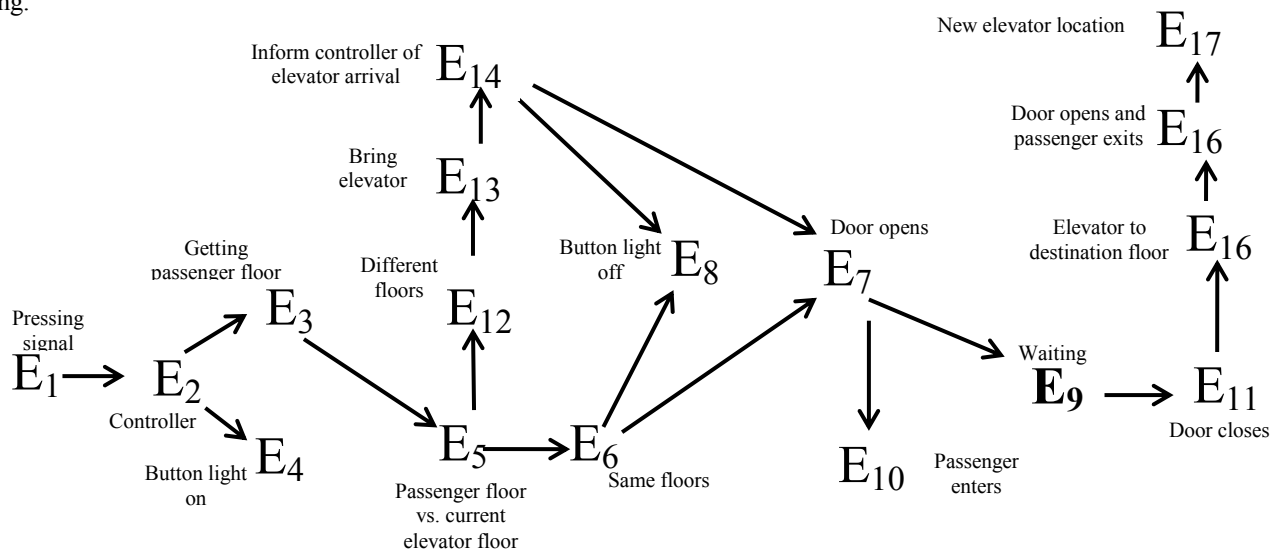

**Fig. 24. The behavioral model of the system of equations.**

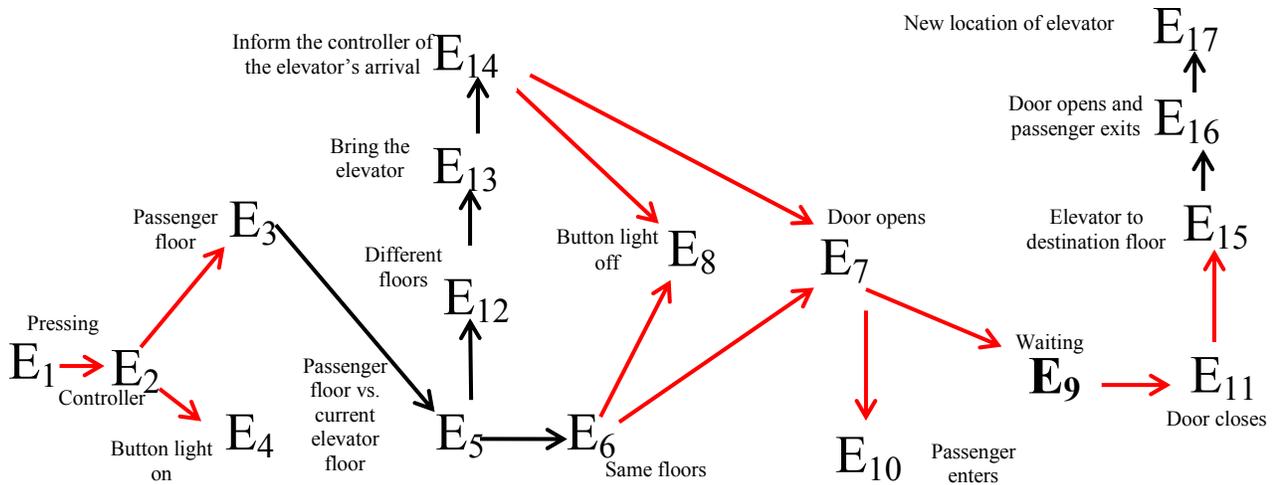

**Fig. 25. The causal links in the behavioral model of the system of equations.**

## REFERENCES

[1] Guizzardi G., Herre H., Wagner G. "Towards Ontological Foundations for UML Conceptual Models", In: Meersman R., Tari Z. (eds) On the Move to Meaningful Internet Systems, pp. 1100–1117, 2002, Lecture Notes in Computer Science, vol 2519. Springer, Berlin, Heidelberg.

[2] J. Pearl, "Theoretical impediments to machine learning with seven sparks from the causal revolution," arXiv preprint, 2018, arXiv:1801.04016.

[3] C. Chendrayan, "A general theory of causality," March 21, 2010, SSRN: https://ssrn.com/abstract=1576102

[4] H. Beebee, "Causation," in The Bloomsbury Companion to Analytic Philosophy (Continuum Companions), B. Dainton and H. Robinson, Eds. London: Bloomsbury Publishing PLC, 2014.

[5] J. Pearl, The Art and Science of Cause and Effect, Epilogue, Causality: Models, Reasoning, and Inference. New York: Cambridge University Press, 2009, pp. 401–428.

[6] R. Guo, L. Cheng, J. Li, P. Richard Hahn, and H. Liu, "A survey of learning causality with data: Problems and methods," ACM Trans. Web., vol 9, no. 4, March 2010.

[7] M. Hulswit, "A short history of 'causation,'" in Cause to Causation: A Peircean Perspective. Dordrecht: Kluwer Publishers, 2002, http://see.library.utoronto.ca/SEED/Vol4-3/Hulswit.pdf

[8] Stanford Encyclopedia of Philosophy, "Counterfactual theories of causation," Oct 29, 2019.

[9] D. Lewis, "Causation," J. Philos., vol. 70, pp. 556–67, 1973.

[10] D. H. Mellor, The Facts of Causation. New York: Routledge, 1995.

[11] C. Khoo, S. Chan, & Y. Niu, "The many facets of the cause-effect relation," in The Semantics of Relationships: An Interdisciplinary Perspective, R. Green, C. A. Bean, and S. H. Myaeng, Eds. CITY: PUBLISHER, DATE, pp. 51–70.

[12] P. Terenziani and P. Torasso, "Time, action-types, and causation: An integrated analysis," Comput. Intell., vol. 11, no. 3, pp. 529–552, 1995.

[13] M. Heidegger, "The thing," in Poetry, Language, Thought, A. Hofstadter, Trans. New York: Harper and Row, 1975, pp. 161–184.

[14] S. Al-Fedaghi, "Modeling the realization and execution of functions and functional requirements," Int. J. Comput. Sci. Inf. Secur., vol. 18, no. 3, March 2020.

[15] S. Al-Fedaghi, "Computer science approach to philosophy: Schematizing Whitehead's processes," Int. J. Adv. Comput. Sci. Appl., vol. 7, no. 11, 2016.

[16] S. Al-Fedaghi and H. Alnasser, "Modeling network security: Case study of email system," Int. J. Adv. Comput. Sci. Appl., vol. 11, no. 3, 2020.

[17] S. Al-Fedaghi, "Thing-oriented learning: Application to mathematical objects," 19th IEEE Int. Conf. Comput. Sci. Eng., Paris, pp. 464–467, Aug. 24–26, 2016.

[18] S. Al-Fedaghi, "Thing/machines (thimacs) applied to structural description in software engineering," Int. J. Comput. Sci. Inf. Sec., vol. 17, no. 8, August 2019.

[19] S. Al-Fedaghi, "Five generic processes for behaviour description in software engineering," Int. J. Comput. Sci. Inf. Sec., vol. 17, no. 7, pp. 120–131, July 2019.

[20] S. Al-Fedaghi, "Toward maximum grip process modeling in software engineering," Int. J. Comput. Sci. Inf. Sec., vol. 17, no. 6, June 2019.

[21] S. Al-Fedaghi, "Thinging as a way of modeling in poiesis: Applications in software engineering," Int. J. Comput. Sci. Inf. Sec., vol. 17, no. 11, November 2019.

[22] S. Al-Fedaghi, "Thinging vs objectfying in software engineering," Int. J. Comput. Sci. Inf. Sec., vol. 16, no. 10, July 2018.

[23] S. Al-Fedaghi, "Thinging for software engineers," Int. J. Comput. Sci. Inf. Sec., vol. 16, no. 7, pp. 21–29, July 2018.

[24] H. G. Steiner, "Theory of mathematics education: An introduction," For Learning Mathematics., vol. 5, no. 2, pp. 11–17, 1985.

[25] A. Sfard, "On the dual nature of mathematical conceptions: Reflections on processes and objects as different sides of the same coin," Educ. Studies in Mathematics, vol. 22, no. 1, pp. 1–36, 1991.

[26] E. Boccardi, "Recent trends in the philosophy of time: An introduction to time and reality i," Rev. Int. Fil. Campinas, vol. 39, no. 4, pp. 5–34, 2016. http://www.scielo.br/pdf/man/v39n4/2317-630X-man-39-04-00005.pdf.

[27] AUTHOR, "Backward causation: Counterfactual theories of causation," in Stanford Encyclopedia of Philosophy. CITY: PUBLISHER, Nov 18, 2015.

[28] A. M. Hoss, "Ontology-based methodology for error detection in software design," PhD dissertation, Department of Computer Science, Louisiana State University, August 2006.